\documentclass[conference]{IEEEtran}
\IEEEoverridecommandlockouts
\usepackage{graphicx}
\usepackage{threeparttable}
\usepackage{multirow}
\usepackage{booktabs}
\usepackage{hyperref}
\def\BibTeX{{\rm B\kern-.05em{\sc i\kern-.025em b}\kern-.08em
    T\kern-.1667em\lower.7ex\hbox{E}\kern-.125emX}}
\begin{document}

\title{DeepC2: AI-powered Covert Command and Control on OSNs}

\author{

\IEEEauthorblockN{Zhi Wang\IEEEauthorrefmark{1}\IEEEauthorrefmark{2},
Chaoge Liu\IEEEauthorrefmark{1}\IEEEauthorrefmark{2},
Xiang Cui\IEEEauthorrefmark{3}}
\IEEEauthorblockA{\IEEEauthorrefmark{1}Institute of Information Engineering, Chinese Academy of Sciences, Beijing, China}
\IEEEauthorblockA{\IEEEauthorrefmark{2}School of Cyber Security, University of Chinese Academy of Sciences, Beijing, China}
\IEEEauthorblockA{\IEEEauthorrefmark{3}Cyberspace Institute of Advanced Technology, Guangzhou University, Guangzhou, China}
\IEEEauthorblockA{wangzhi@iie.ac.cn liuchaoge@iie.ac.cn cuixiang@gzhu.edu.cn}
}

\author{
  \IEEEauthorblockN{Zhi Wang, Chaoge Liu, Xiang Cui, Jie Yin, Jiaxi Liu, Di Wu, Qixu Liu}
}

\maketitle
\thispagestyle{plain}
\pagestyle{plain}
\begin{abstract}
Command and control (C\&C) is important in an attack. It transfers commands from the attacker to the malware in the compromised hosts. Currently, some attackers use online social networks (OSNs) in C\&C tasks. There are two main problems in the C\&C on OSNs. First, the process for the malware to find the attacker is reversible. If the malware sample is analyzed by the defender, the attacker would be exposed before publishing the commands. Second, the commands in plain or encrypted form are regarded as abnormal contents by OSNs, which would raise anomalies and trigger restrictions on the attacker. The defender can limit the attacker once it is exposed. In this work, we propose DeepC2, an AI-powered C\&C on OSNs, to solve these problems. For the reversible hard-coding, the malware finds the attacker using a neural network model. The attacker's avatars are converted into a batch of feature vectors, and the defender cannot recover the avatars in advance using the model and the feature vectors. To solve the abnormal contents on OSNs, hash collision and text data augmentation are used to embed commands into normal contents. The experiment on Twitter shows that command-embedded tweets can be generated efficiently. The malware can find the attacker covertly on OSNs. Security analysis shows it is hard to recover the attacker's identifiers in advance.
\end{abstract}


\section{Introduction}\label{sec:intro}
Command and control (C\&C) plays an essential role in an attack. It is widely used in Advanced Persistent Threat (APT), ransomware, or botnet scenarios. In a C\&C system, an attacker needs to send commands to the compromised hosts via a C\&C channel~\cite{Bailey09}. The hosts can be common computing devices such as PCs, servers, routers, and cameras, which have some vulnerabilities and can be infected by malware. They try to get and execute the commands from the attacker and carry out attack tasks such as DDoS, spam, crypto-mining, and data exfiltration. The major feature of a C\&C system is that it has a one-to-many C\&C channel, which receives commands from the attacker and forwards them to the compromised hosts. The process for the malware getting the commands is called addressing. C\&C channel is a vital component in a C\&C system. The attacker needs to keep the C\&C channel robust and block-resistant to maintain the communication with the malware.

In recent years, the attackers have begun to utilize online web services~\cite{Yin18}, such as online social networks (OSNs), cloud drives, and online clipboards, to build the C\&C channel.
For example (as shown in Table~\ref{tab:ccs}), Hammertoss (APT-29)~\cite{FireEye15} used Twitter and GitHub to publish commands and hide communication traces. HeroRat~\cite{Stefanko18} used Telegram for C\&C communication on Android devices. Turla~\cite{Faou20} utilized Gmail to receive commands and exfiltrate information to the operators.

OSNs have some features to build a good C\&C channel. 
It is nearly impossible for OSNs to go offline, and users can access OSNs anytime with a networked device. 
Then, visiting OSNs is allowed by most anti-virus software, and it ensures the availability of the commands.
As many people use the OSNs, the attacker's accounts can hide among ordinary users.
Also, it is easy to limit the accounts but not easy to shut down the OSNs. With the help of dynamic addressing, the malware can obtain commands from multiple accounts.

\begin{table}
  \caption{Carriers for building C\&C channels}
  \label{tab:ccs}
	  \begin{tabular}{ccccc}
	    \toprule
	    \textbf{Year}&\textbf{Name}&\textbf{Platform}&\textbf{Identity}&\textbf{CMD}\\
	    \midrule
	    2009&upd4t3~\cite{upd4t309}&Twitter&ID&Base64\\
	    2012&-~\cite{singh2012social}&Twitter&Token&Plain\\
	    2014&Garybot~\cite{sebastian2014framework}&Twitter&Token&Encrypted\\
	    2015&Hammertoss~\cite{FireEye15}&Twitter&DGA&Plain\\
	    2015&MiniDuke~\cite{F-Secure15}&Twitter&DGA&Base64\\
	    2015&-~\cite{Pantic15}&Twitter&DGA&Tweets\\
	    2017&ROKRAT~\cite{ROKRAT17}&Twitter&ID, Token&Plain\\
	    2017&PlugX~\cite{PlugX17}&Pastebin&URL&XOR\\
	    2018&Comnie~\cite{Comnie18}&GitHub&URL&Base64, RC4\\
	    2019&DarkHydrus~\cite{DarkHydrus19}&GoogleDrive&URL&Plain\\
	    2020&Turla~\cite{Faou20}&Gmail&Token&Encrypted\\
	  \bottomrule
	\end{tabular}
\end{table}

\begin{figure}
    \centering
    \includegraphics{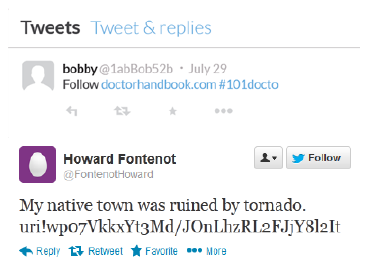}
    \caption{Commands posted by Hammertoss and MiniDuke}
    \label{fig:apt}
\end{figure}

However, there are two main problems with building C\&C channels on OSNs. First, the attacker's identifiers are reversible and predictable, which will cause the C\&C channel to shut down before use. To help the malware addressing, the attacker's identifiers, i.e., ids, links, tokens, and DGAs (Domain Generation Algorithms), have to be hard-coded into the malware. Once the malware is analyzed by defenders, the reversible hard-coding will expose the C\&C channel, and the attacker's accounts can be calculated in advance. Second, the commands published on OSNs are abnormal and will also expose the attacker's accounts.
In most cases, commands are published in plain, encoded, or encrypted forms, as shown in Fig.~\ref{fig:apt}. They are regarded as abnormal contents on OSNs. They will expose C\&C activities and raise anomalies, triggering restrictions on the attacker's accounts and interrupting the C\&C activities. If the OSNs block the accounts, it is difficult for the malware to retrieve new commands.

In this paper, from the attacker's perspective, we use AI technology to overcome the above two problems and propose an AI-powered OSN C\&C channel called DeepC2.
The main idea of DeepC2 is as follows. To overcome the first problem, a neural network model is used for addressing. The malware finds the attacker's accounts through the feature vectors, which are extracted from the attacker's avatars by a neural network model. As the neural network models are poorly explainable~\cite{xai20}, defenders cannot calculate and predict the avatars and accounts through the model and vectors. To solve the second problem and eliminate the abnormal content, we propose embedding the commands into contextual and readable content (we take Twitter and tweets as examples). To achieve this, the attacker uses data augmentation to generate numerous tweets and uses hash collision to get the command-embedded tweets. In the addressing process, Twitter Trends are used as the rendezvous point. The attacker posts tweets to a trending topic, and the malware finds the attacker under the topic. After addressing, the commands can be parsed from the attacker's tweets.

The contributions of this paper are summarized as follows:

\begin{itemize}
\item We propose a novel covert command and control scenario on OSNs.
\item We introduce neural networks to solve the problem of reversible hard-coding in C\&C addressing. By using feature vectors and a model, it is easy for the malware to find the attacker while hard for defenders to locate the attacker in advance.
\item We propose a method for embedding commands into natural semantic tweets to avoid anomalies caused by abnormal contents on OSNs. 
\item We present experiments on Twitter to demonstrate the feasibility of the proposed methods and analyze their performance and security.
\end{itemize}

\textbf{Ethical Considerations.} The combination of AI and network attacks is an upward trend. We cannot stop the evolution of cyberattacks, but we should draw attention to the defenses in advance. This work aims not to inspire malware authors to write more efficient malware but to motivate security researchers and vendors to find solutions for an emerging threat.
To this end, we intend to provide this work to build a possible scenario to help prevent this kind of attack in advance.

The remainder of this paper is structured as follows. Section~\ref{sec:background} describes relevant backgrounds and related work. Section~\ref{sec:method} presents the methodology for building the covert C\&C channel. Detailed implementations are demonstrated in Section~\ref{sec:imple}. Section~\ref{sec:eva} is the evaluations on the experiments. Section~\ref{sec:discuss} discusses possible countermeasures. Conclusions are summarized in Section~\ref{sec:conclusion}.

\section{Background and Related Work}\label{sec:background}
In this section, we present the background and related work of DeepC2.

\subsection{Command and Control on OSNs}\label{sec:botnet}

In the cases of addressing using OSN platforms, the defenders should find and limit the attacker's accounts in advance. By reverse-engineering a malware sample, the defenders will know the addressing process in detail~\cite{DGA16}. If reversible methods like DGAs or IDs are used, the attacker's accounts can be calculated in advance. The defenders can limit the accounts, making the C\&C channel unusable. The defenders will also know the attacker's accounts and commands by running a sample. However, when they get the commands this way, the malware in the wild also gets the commands, which is a failure from the defense perspective~\cite{pony2021}. Therefore, the key issue in such an attack is to design a block-resistant C\&C channel that even the defenders know the detailed information about the channel, it is hard to get the attacker's identifiers and limit the C\&C \textbf{in advance}.

Some works build C\&C channels on OSNs. 
Stegobot~\cite{Shishir11} uses the images shared by OSN users to build the C\&C channel. The social network is regarded as a peer-to-peer network to connect the malware and the attacker. Information is hiding in images using steganography. However, the attacker's account can be obtained through reverse-engineering.
Sebastian et al.~\cite{sebastian2014framework} proposed to build a covert C\&C channel on Twitter. The commands are encrypted tweets with a keyword, for example, \#walmart AZEF, where \#walmart is the keyword, and AZEF is the command cipher. However, this method also has the problem of abnormal contents on OSNs.
Kwak et al.~\cite{Kwak21} proposed a video steganography-based C\&C channel on Telegram, which can transfer large-sized secret files.
Pantic et al.~\cite{Pantic15} proposed an anomaly-resistant C\&C on Twitter. They used tweet-length as a command character and encoded each symbol in commands into numbers from 1 to 140. They collected tweets at different lengths from Twitter. When publishing a command, they chose tweets at specified lengths and posted them. The malware can get the commands by calculating the lengths of the tweets. However, this method has a low capacity and does not solve the reversible attacker accounts.

\subsection{Easy Data Augmentation}\label{sec:eda}
Data augmentation is a technique to solve the insufficiency of training data.
By applying data augmentation, researchers can enlarge the existing dataset to meet the needs of training works and promote the normalized performances of neural network models.
In this work, the attacker needs to generate numerous tweets for hash collisions. Wei et al.~\cite{Wei19} proposed Easy Data Augmentation (EDA) techniques. They used Synonym Replacement (SR), Random Insertion (RI), Random Swap (RS), and Random Deletion (RD) to generate sentences with similar meanings to the given sentences. Examples of EDA are shown in Table~\ref{tab:eda} with an original sentence from Twitter~\cite{MITREattack20}.

The augmented sentences may not be grammatically and syntactically correct and may vary in meaning. However, due to differences in language, culture, and education, there are many grammatically incorrect tweets on Twitter. The Internet is diverse and inclusive. The attacker should ensure that the tweets have semantics but do not need them to be ``correct''.

\begin{table}[]
\caption{Sentences generated by EDA}
\label{tab:eda}
\begin{tabular}{cp{6cm}}
\toprule
\textbf{Operation} & \textbf{Sentence}                                                         \\ \midrule
None      & Our TAXII server is going to be taking a short nap at 11am ET today for an update. \\ \midrule
SR        & Our TAXII server is \textbf{\textit{endure}} to be taking a short nap at 11am ET today for an update.          \\ \midrule
RI        & Our TAXII server is going to be taking a short nap at 11am \textbf{\textit{cat sleep}} ET today for an update. \\ \midrule
RS        & Our \textbf{\textit{short}} server is going to be taking a \textbf{\textit{TAXII}} nap at 11am ET today for an update. \\ \midrule
RD        & Our server is to be taking a short nap at 11am ET today for an update.            \\ \bottomrule
\multicolumn{2}{l}{SR: synonym replacement. RI: random insertion. }\\
\multicolumn{2}{l}{RS: random swap. RD: random deletion.}
\end{tabular}
\end{table}

\begin{table*}
    \centering
    \caption{Contents access restrictions of Alexa top OSN sites}
    \resizebox{0.9\textwidth}{!}{
    \begin{tabular}{|c|c|c|c|c|c|c|c|c|c|c|c|c|c|c|c|}
      \hline
      \multirow{2}*{\textbf{OSN}}&\multicolumn{3}{|c|}{\textbf{Profiles}}&\multicolumn{3}{|c|}{\textbf{Posts}}&\multicolumn{3}{|c|}{\textbf{Comments}}&\multicolumn{3}{|c|}{\textbf{Pictures}}&\multicolumn{3}{|c|}{\textbf{Trends}}\\ \cline{2-16}
      &Login&Area&All&Login&Area&All&Login&Post&All&Compress&Watermark&Resize&Login&Area&All\\
      \hline
      facebook.com&N&N&C&N&N&Y&N&Y&Y&Y&N&N&-$^{\mathrm{*}}$&-&-\\
      twitter.com&N&N&Y&N&N&C&N&Y&C&Y&N&N&N&R&Y\\
      instagram.com&Y&N&Y&Y&N&C&Y&Y&Y&Y&N&O&Y&N&Y\\
      weibo.com&N&N&C&N&N&C&N&Y&Y&Y&Y&Y&N&N&Y\\
      tumblr.com&N&N&Y&N&N&Y&N&Y&C&Y&N&Y&N&N&Y\\
      imgur.com&N&N&Y&N&N&Y&N&Y&Y&Y&N&N&N&N&Y\\
      pixnet.net&N&N&C&N&N&Y&N&Y&Y&Y&N&Y&N&N&Y\\
      pinterest.com&Y&N&Y&Y&N&Y&Y&Y&Y&Y&N&N&R&Y&N\\
      \hline
      \multicolumn{5}{l}{$^{\mathrm{*}}$ facebook.com does not provide trends.}\\
      \multicolumn{12}{l}{Login – Login to view, Area – Area restrictions, All – All contents is available, Post – Login to post}\\
      \multicolumn{12}{l}{Y – Yes, N – No, O – Occasionally, C – Customized by user, R – Restrictions that can be bypassed.}
    \end{tabular}
    }
    \label{tab:osn}
\end{table*}

\subsection{Online Social Networks}\label{sec:osns}
People share content in different forms on OSNs.
Due to different policies and user privacy settings, content access permissions vary on different OSNs. Some OSNs' contents are limited only to authenticated users, while some have no restrictions that everyone can access all contents in the platform. Table~\ref{tab:osn} shows the content access restrictions of Alexa top OSN sites. Attackers can utilize the nonrestricted parts to convey customized information to others (including bots). In this work, to demonstrate the feasibility of the methods, we choose Twitter to build the C\&C channel. The commands are embedded in tweets and posted by the botmaster. 

\subsection{AI-powered Attacks}
This work provides a new scenario on the malicious use of AI. The combination of AI and network attacks is an upward trend.
For covert communication, Rigaki et al.~\cite{RigakiG18} proposed using GAN to mimic Facebook chat traffic to make C\&C communication undetectable. StegoNet~\cite{Liu20stego} and EvilModel~\cite{Wang2021EvilModel, Wang2022EvilModel2} hide malware in the neural network models to deliver malicious payloads covertly. The model parameters are replaced by the malicious payloads. 
DeepLocker~\cite{Kirat18} can carry out targeted attacks stealthily. DeepLocker trains the target attributes into an AI model and uses the model's outputs as a symmetric key to encrypt the malicious payload. Target detection is conducted by the AI model. When the input attributes match the target attributes, the secret key will be derived from the model to decrypt the payload and launch attacks on the target. 
MalGAN~\cite{HuT17} generates adversarial malware that can bypass machine learning-based detection models. A generative network is trained to minimize the malicious probabilities of the generated adversarial examples predicted by the detector. More evasion methods~\cite{EvadePE18}\cite{Wang21} were proposed after MalGAN.

\section{Methodology}\label{sec:method}

This section introduces methodologies for building a covert C\&C channel on OSNs.

\subsection{Threat Model}
In this work, the C\&C channel is built from the attackers' perspective. This work has three prominent roles: attackers, OSNs, and defenders.
\textbf{Attackers}. We consider adversaries to be attackers capable of neural network and artificial intelligence technics and have the ability to use various system vulnerabilities to get into a system.
\textbf{OSNs}. OSNs have the ability to limit the abnormal content and accounts based on their term of services. They can also actively detect autonomous and abnormal behaviors and limit the specious accounts according to their regulations.
\textbf{Defenders}. We consider defenders to be third-party unrelated to attackers and OSNs. Defenders have access to the vectors from the prepared pictures and the structure, weights, implementation, and other detailed information of the neural network model. Defenders also have the ability to reverse engineer the malware sample to obtain its detailed implementation.

\subsection{Approach Overview}\label{sec:overview}

\subsubsection{Overall Workflow} 
\begin{figure}
\centering
\includegraphics[width=\linewidth]{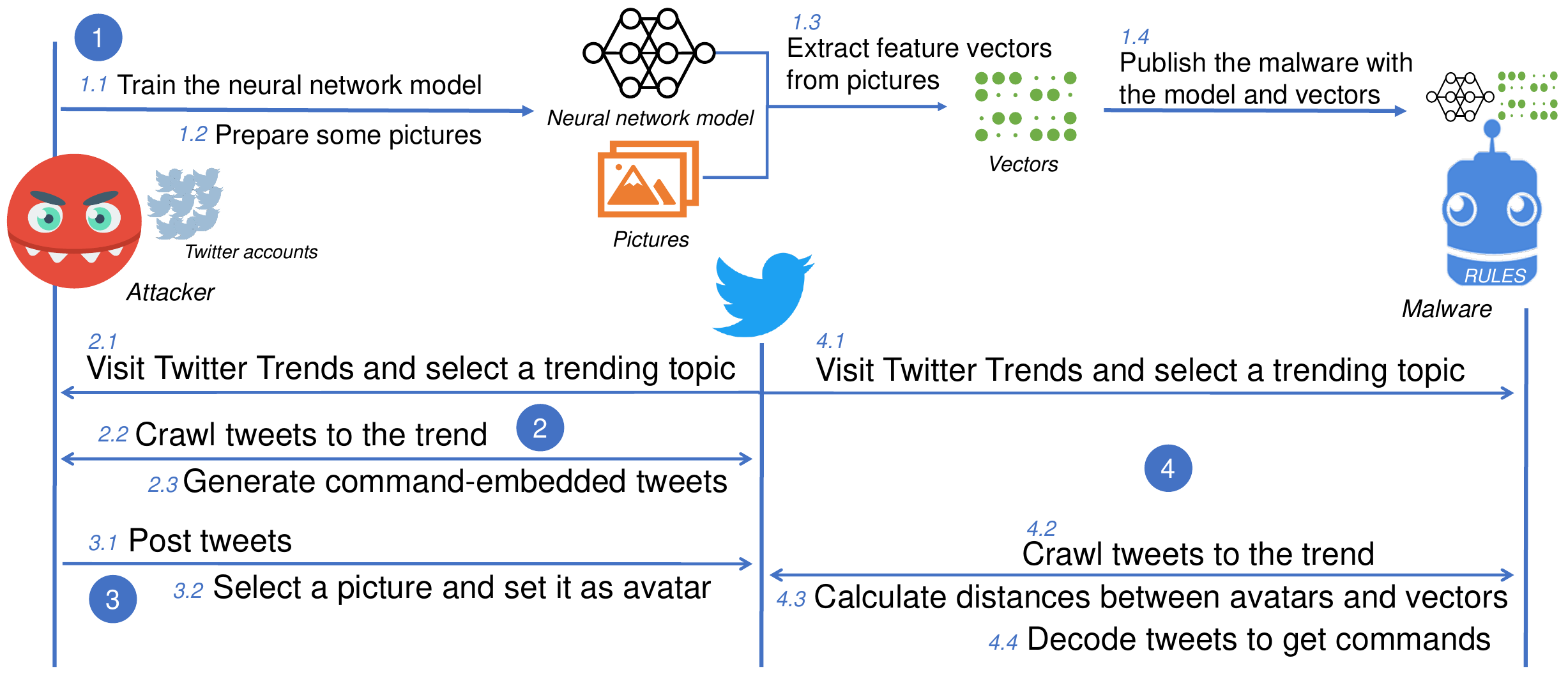}
\caption{Overall workflow of DeepC2}
\label{fig:workflow}
\end{figure}

We take Twitter as the OSN platform to demonstrate the method. The main workflow of DeepC2 contains four steps and is shown in Fig.~\ref{fig:workflow}.

\begin{figure}[!t]
\centering
\includegraphics[width=0.6\linewidth]{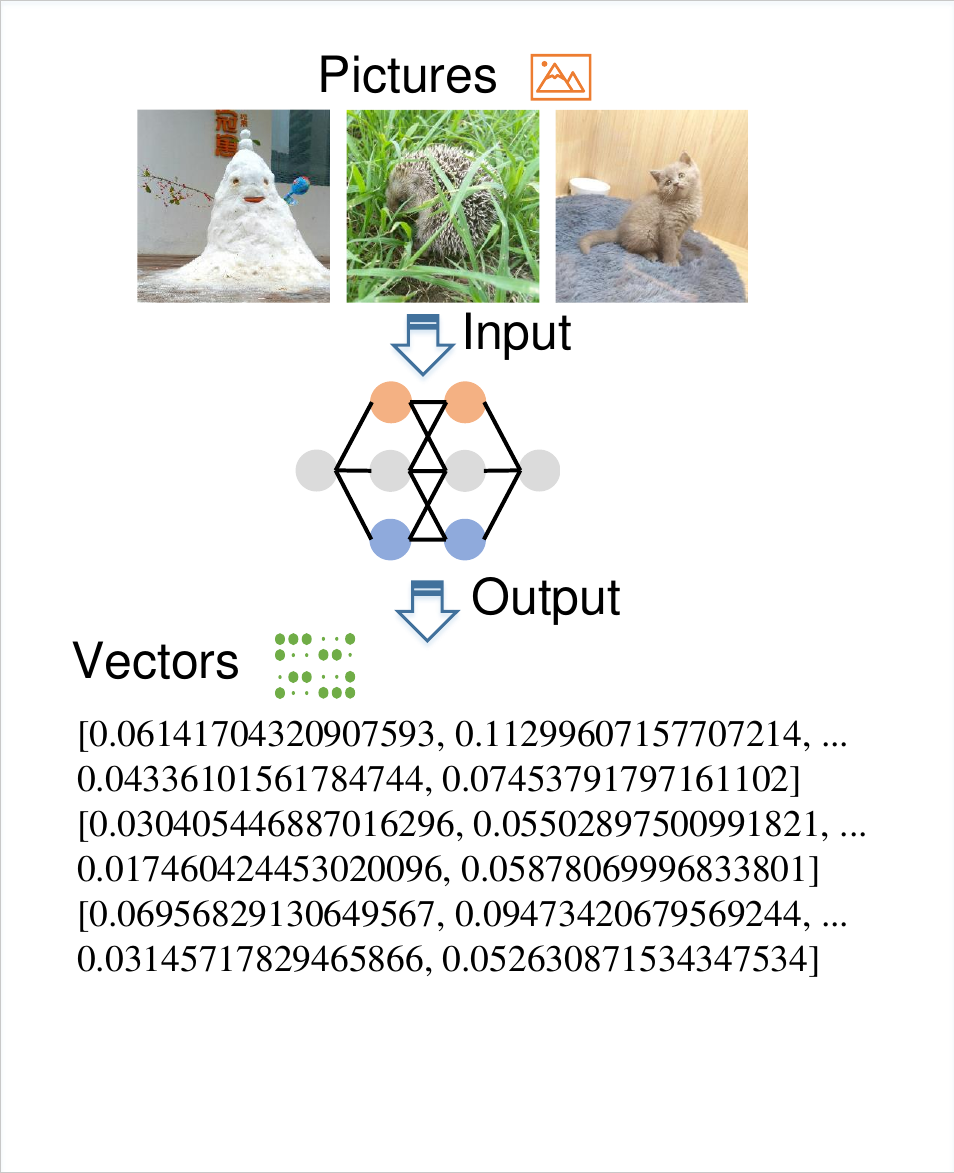}
\caption{Extract feature vectors}
\label{fig:pic2vec}
\end{figure}

(1) The attacker trains a neural network model with some pictures that may be selected as Twitter account avatars in subsequent steps. Then, the attacker extracts the feature vectors of these pictures using the trained model (see Fig.~\ref{fig:pic2vec}), embeds the model and the feature vectors into the malware, and publishes the malware to the wild.

(2) The attacker visits Twitter Trends, selects a topic according to the pre-defined rules, and then generates command-embedded tweets based on the selected trending topic.

(3) The attacker selects a picture as its Twitter account avatar and publishes the commands-embedded tweets using the account in the selected topic.

(4) The compromised hosts infected with the attacker's malware visit Twitter Trends periodically, select a topic synchronously according to the pre-defined rules, and then crawl the tweets and tweeters' avatars in the selected topic to find the attacker's account. The compromised hosts calculate the distances between the crawled avatars and the built-in feature vectors. If a distance is below a threshold, it is considered that the attacker's account is found. The commands can be obtained from the tweet.

The workflow mainly contains two key parts: dynamic addressing and command embedding. The dynamic addressing guides the compromised host to find the attacker and get the command successfully. Rather than reversible methods, it uses a neural network model, picture feature vectors, and Twitter topics to avoid the attacker's identifiers being exposed and predicted.
The command embedding uses hash collision and EDA to embed commands into natural semantic tweets, thus avoiding anomalies caused by abnormal contents on OSNs.
Next, we will describe these two parts in detail.

\subsection{Dynamic Addressing}

The dynamic addressing includes three main elements: the neural network model, avatars \& feature vectors, and Twitter trending topics. In the following section, we will describe the elements from three aspects: What (is the element), How (to use), and Why (is the element).

\begin{figure}[!t]
\centering
\includegraphics[width=0.8\linewidth]{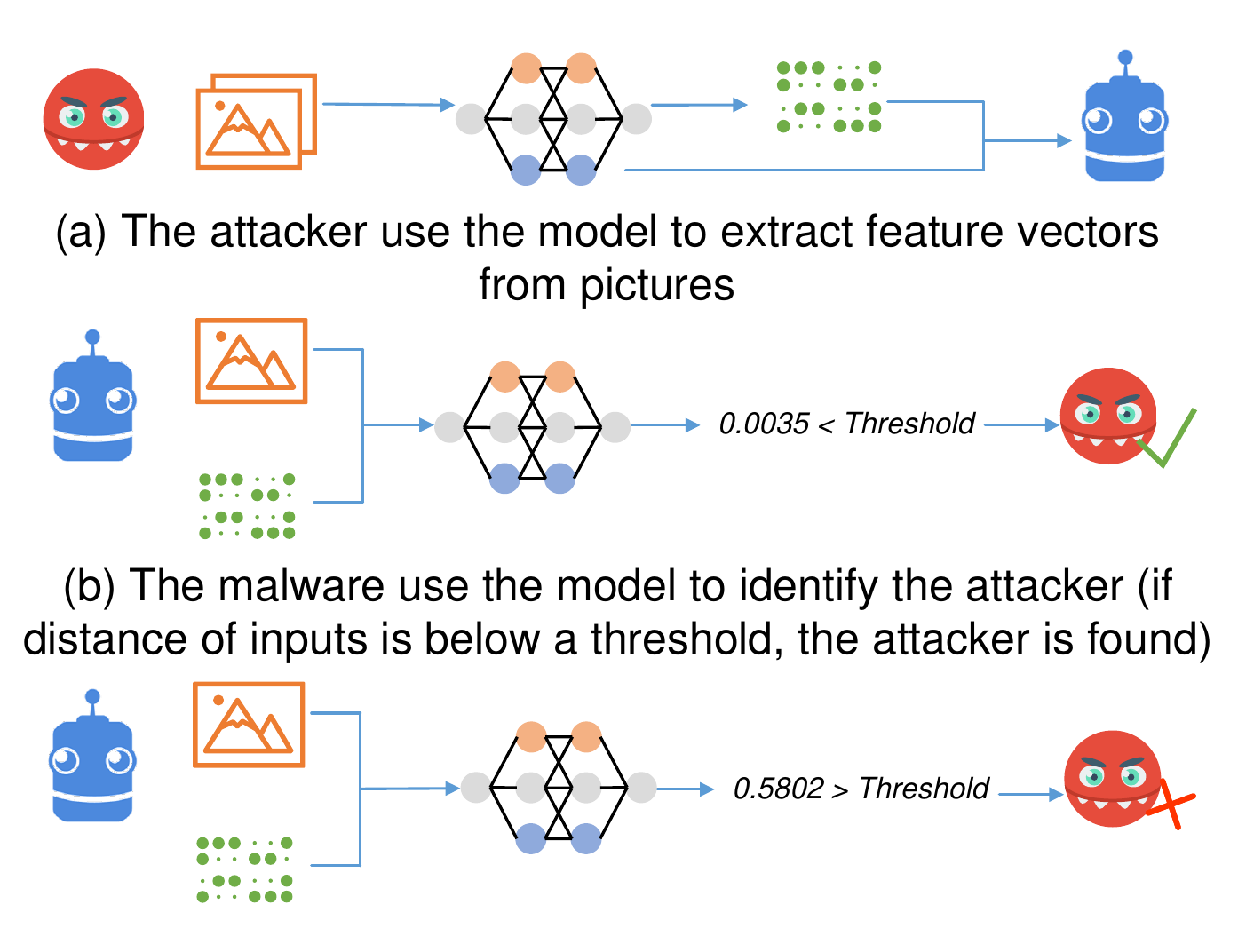}
\caption{Neural network model use}
\label{fig:usage}
\end{figure}

\subsubsection{Neural Network Model}
\textbf{What}. The attacker needs a neural network model to extract feature vectors, and the malware needs the model to identify the attacker's accounts. Due to the limited resources in the host devices, the attacker cannot use the big-sized pre-trained models like VGGs, AlexNets, and Inception. Therefore, the attacker needs to build and train a model itself.

\textbf{How}. The model is used differently for attackers and malware (see Fig.~\ref{fig:usage}). For the attacker, the model is used to extract feature vectors from avatars. The attacker feeds the model with a batch of pictures, and the model outputs a batch of vectors that represent the pictures. The feature vectors and the model are published with the malware.
For the malware, the model calculates the distances between avatars from Twitter users and the vectors to identify attackers. A selected vector and a crawled avatar are fed into the model, and then the model outputs the distance of the inputs.

\textbf{Why}. Using neural network models has the following advantages:
1) It is not easy to reverse the neural network model. Convolution is a lossy process. Combined with some intentionally introduced losses, it is hard to calculate the attacker's identifiers in advance.
2) The neural network models are fault-tolerance that similar inputs will generate similar outputs.
3) The neural network trained with gradient descent has good generalization ability~\cite{google_nnmodel22}. It can help the malware identify attackers accurately and does not mistakenly identify someone else as the attacker.

Converting an image to a vector is similar to image-hashing~\cite{ImageHash2021}. In image-hashing, similar inputs have similar output hashes. However, image-hashing is not suitable for this work. There are two types of image-hashing methods. The non-neural network-based image-hashing methods are reversible, and defenders can build images that produce similar vectors according to the given hashes. For neural network-based methods~\cite{ImageHashAAAI}, the learning tasks are more complex than DeepC2. The cryptographic hash algorithms are also unsuitable because they are sensitive to changes. As pictures uploaded to OSNs are compressed or resized, avatars are different from the original images, which will cause hashes to change due to the avalanche effect~\cite{sbox1985}. Therefore, the neural network model is suitable for this task.

\subsubsection{Avatars and Vectors}
\textbf{What}. The feature vectors are the abstract expression of the attacker's Twitter avatars. They are generated by the model and represent the attacker's avatars. A feature vector sample is shown in Fig.~\ref{fig:pic2vec}. Each vector contains a group of floating-point numbers, and the amount is determined by the model.

\textbf{How}. As stated above, the attacker gets the feature vectors from the model. The malware puts a feature vector and an avatar in the model and gets the distance between the inputs. The model will convert the avatar into another vector and calculate the distance. To prevent replay and enhance security, it is recommended that each avatar and vector be used only once. The attacker will change the current account and avatar when a command is published, and the malware will also delete the used vectors. The malware can get updates to the vectors and model and exploits from the C\&C server. The malware can carry at least one vector in design when published. Due to various situations, the malware may not be able to run on time. To ensure that the compromised hosts can go online as expected, it is suggested that the malware is published with more vectors.

\textbf{Why}. The reason for using the feature vectors are as follows: 1) They are the natural output of the model and are easy to get for both the attacker and the malware. 2) It's difficult to reverse the vector generation process to get another picture that can produce a similar vector. 3) The vectors are distributed in a continuous interval (see Sec.~\ref{sec:sec_ana}), and each position has a large value space, which ensures the security of the C\&C channel.

\subsubsection{Twitter Trends}\label{sec:trends}
\textbf{What}. Twitter Trends contains hot topics that have had many discussions in the past 24 hours. Usually, each topic contains 1 to 3 keyword(s). Twitter Trends is updated every 5 minutes.

\textbf{How}. The attacker defines a set of rules to select the trending topics, and the malware selects the topics synchronously with the attacker. In this work, we use Twitter API to get the trending topics. Twitter Trends API returns the top 50 topics in a chosen area specified by a location ID (WOEID, where on Earth ID). There are detailed tweet volumes if the volume exceeds 10 K over the past 24 hours. In the experiments, we obtained trends from Johannesburg, South Africa and selected the last topic above 10 K discussions from the returned trends. The test servers in different regions fetched the same topics with the WOEID. 
Twitter API can be abused by attackers. However, attackers have more choices in real scenarios. They can utilize third parties that provide Twitter content queries, including tweets, trends, and user profiles. Alternatively, they can also write their implementations that use raw HTTP requests to obtain the content. 

\textbf{Why}. Using Twitter Trends have the following advantages: 1) Twitter Trends provides a rendezvous point for the malware to find attackers among Twitter users; 2) Twitter Trends changes with the tweet volume under different topics and is updated every 5 minutes, which is difficult to predict; 3) Since normal users also discuss different topics, attackers can hide among them.
Therefore, we use Twitter Trends for DeepC2.

\subsection{Command Embedding}\label{sec:gentweets}

The attacker uses hash collision and easy data augmentation (EDA) to generate commands-embedded tweets. In this work, we take publishing an IP address as an example to illustrate the process of publishing commands. If IP addresses can be published, other messages like domains, shortening-URL IDs, or online-clipboard IDs can also be published in the same way.

\begin{figure}
\centering
\includegraphics[scale=0.8]{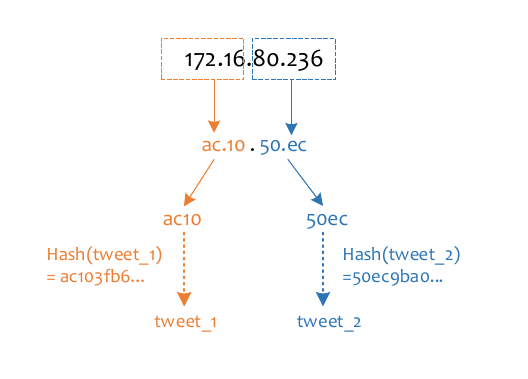}
\caption{Hash collision}
\label{fig:hash}
\end{figure}

(1) \textbf{Hash Collision}. To convey an IP address to the malware through tweets, the attacker splits the IP address into two parts first, as shown in Fig.~\ref{fig:hash}. Each part is expressed in hexadecimal. For each tweet, the attacker calculates its hash and compares whether the first 16 bits of the hash are identical to one IP part. If two parts of an IP address collide, a successful hash collision occurs. The attacker posts the collided tweets in order. When addressing, the malware can get the IP address by calculating the hashes of tweets posted by the attacker and concatenating the first 2 bytes of hashes. In this way, 16 bits can be conveyed in one tweet.

(2) \textbf{Tweets Generation}. To perform a successful collision, the attacker needs numerous tweets. The new tweets are generated using EDA. After selecting a topic, the attacker crawls the trending tweets to generate more sentences. We crawled 1 K tweets for each selected trend in the experiments. Before generating new sentences using EDA, we cleaned the tweets first. As there are word deletions and swaps during augmentation, if a tweet is too short, the generated sentences may not contain the trending words. Thus, we filtered out tweets with less than ten words. Additionally, there were retweeted tweets that did not contain the trending words, so we filtered them out and retained only the original tweets. Then, we removed unnecessary chars like emojis, links, tabs, and line breaks in each tweet. Duplicate tweets were removed at last. Normally there were 400 to 900 tweets left. We used EDA to generate 50 sentences for each remaining tweet. It will get us 20 K to 45 K new sentences. It is still insufficient for a hash collision. We converted all sentences to the upper case and added punctuation (``.'', ``..'', ``...'', ``!'', ``!!'' and ``!!!'') at the end of each sentence. It resulted in 140 K to more than 300 K sentences in total, which greatly increased the success rate for a hash collision (see Sec.~\ref{sec:hashexp}).

It's not recommended to convey a whole IP address in a tweet because it needs too many tweets to perform a successful collision. Two 16 bits will reduce the calculation greatly. Also, it is not deterministic for a successful hash collision. If a collision fails, the attacker can crawl more tweets or add more noise to the sentences. The attacker needs to post the two final tweets in order so that the malware can correctly recover the IP address.

There may be different situations where the compromised hosts cannot go online as expected, and the defenders can put on a saved avatar and post tweets with fake commands. In case of this happening, authentication like a digital signature with asymmetric key pairs is recommended to ensure a secure communication.

\section{Implementation}\label{sec:imple}
In this section, we demonstrate the proposed convert C\&C channel is feasible by presenting a proof-of-concept experiment on Twitter.



\subsection{Siamese Neural Network}

\begin{figure}[!t]
	    \centering
	    \includegraphics[width=\linewidth]{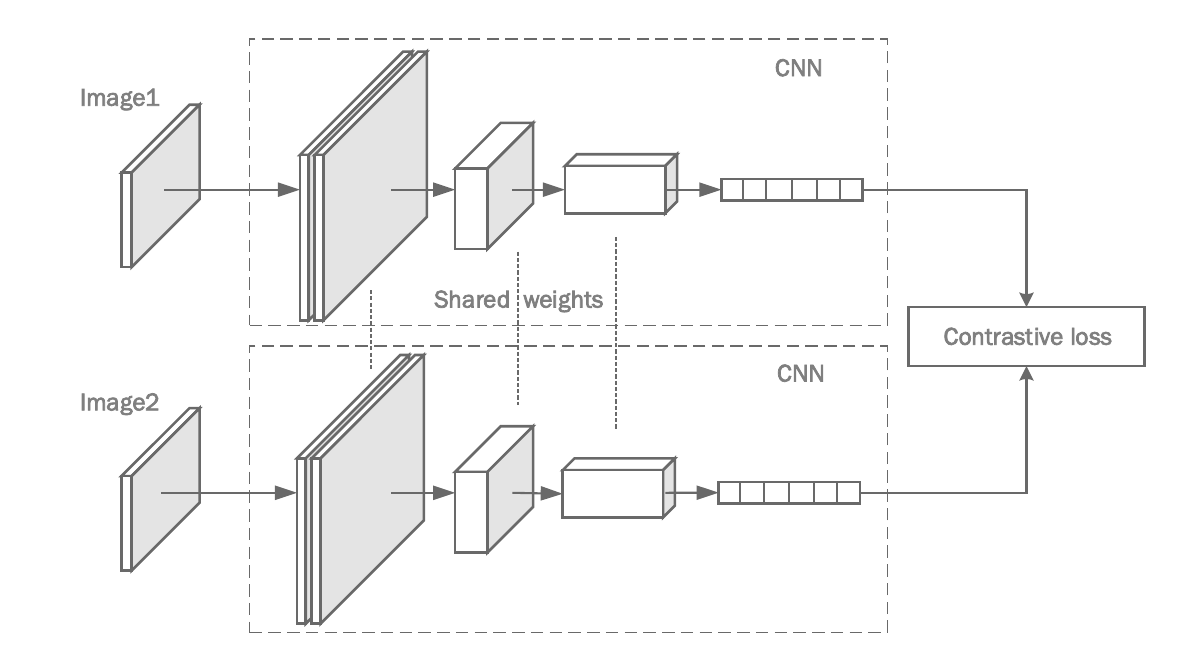}
	    \caption{Architecture of Siamese neural network}
	    \label{fig:snn}
\end{figure}

(1) \textbf{Architecture}. The Siamese Neural Network (SNN)~\cite{snn94} is effective in measuring the similarity between two inputs. The two inputs accepted by SNN will feed into two identical neural networks to generate two outputs. Like ``Siamese'' twins sharing the same organs, the identical neural networks share the same architecture and weights. The similarity between two inputs can be measured by calculating the distance between two outputs. We use Euclidean distance in this work.
Fig.~\ref{fig:snn} shows the architecture of the SNN. In this work, the two identical neural networks are CNNs~\cite{LuCun89}. 
It contains four convolutional layers and three fully connected layers. It accepts a 3-channel 128-pixel image as the input and generates 128 outputs to make up a feature vector.


The contrastive loss function~\cite{LeCun06} is used during the training. For two image inputs of the CNNs, $Y$ is a binary label assigned to the pair, where $Y=0$ represents the images being similar, and $Y=1$ means that the images are different. $G_1$ and $G_2$ are two vectors generated by identical CNNs. Let $D_w=\|G_1-G_2 \|$ be the Euclidean distance between the vectors, $w$ be the weights of the network, and $m>0$ be a margin (radius around $G$). The loss function is:
\[L=(1-Y) \frac {1} {2} (D_w )^2 + Y \frac {1} {2} (\max (0, m-D_w) )^2\]

(2) \textbf{Training}. The model was implemented with Python 3.6 and PyTorch 1.5. To train the model, we crawled avatars of different sizes from 115,887 Twitter users and randomly selected 19,137 sets of avatars to build the dataset. Twitter provides 4 different sizes of avatars: 48x48, 73x73, 200x200 and 400x400. We randomly chose avatars of size 400x400 to make up input pairs with label 1. Due to the lack of original pictures of the avatars, we used avatars with sizes of 200x200 and 400x400 from the same user to make up input pairs with label 0. The ratio of input pairs marked as 0 and 1 is 1:2.
Based on a preliminary experiment (Appendix~\ref{app:threshold}), the threshold for Euclidean distance was set to 0.02. 

(3) \textbf{Performance}. To test the performance, we conducted the training process several times. The model converged rapidly during training. After 10-20 epochs, 100\% accuracy on the test set was obtained. 
The size of a trained model is 2.42 MB. We used avatars from all 115,887 users to make up the validation set, for a total of 463,544 pairs (115,887 pairs with label 0 and 347,657 pairs with label 1, 1:3 in ratio). Evaluations show that the model reached an accuracy of more than 99.999\%, with only 2-4 mislabeled pairs. Different from traditional machine learning works, we need to avoid hijacking the attacker's accounts, which means mislabeling from \textit{not the same} to \textit{the same} (false positive) is forbidden, while some mislabeling from \textit{the same} to \textit{not the same} (false negative) is allowed. The original labels of the mislabeled pairs were all 0, which means no avatar collision occurred with the trained models. It ensured the security of the attacker's accounts.

\begin{figure}[!t]	    
	    \centering
	    \includegraphics[width=\linewidth]{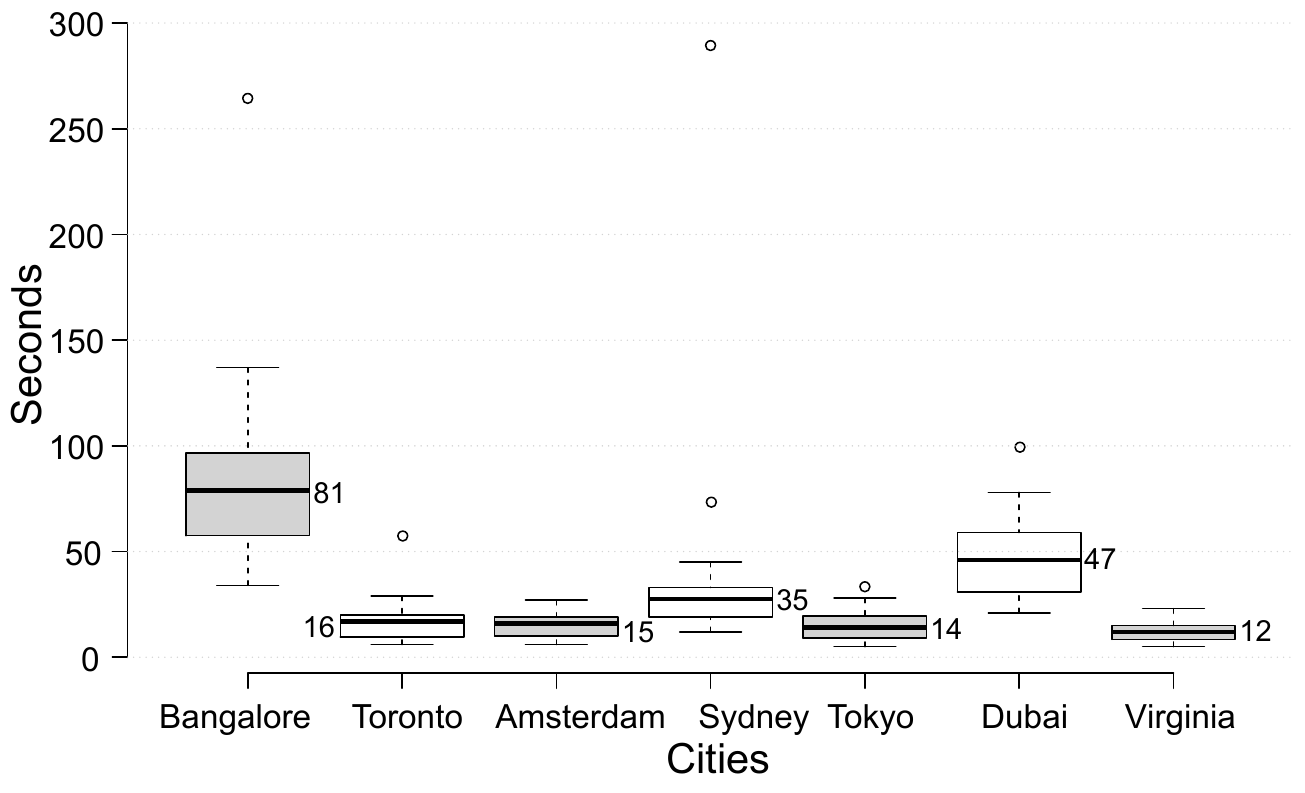}
	    \caption{Time cost for finding attacker}
	    \label{fig:mainres}
\end{figure}

\subsection{Experiments on Twitter}\label{sec:exp}
(1) \textbf{Environments}.
To simulate the compromised hosts worldwide, we used 7 Ubuntu 18.04 x64 virtual servers with 1 GB ROM and 1 vCPU located in Bangalore, Toronto, Amsterdam, Sydney, Tokyo, Dubai, and Virginia. The code for the attacker was run on another virtual server with the same configuration in San Francisco. Both codes for the malware and attacker were implemented with Python 3.6.

(2) \textbf{Commands and Avatars}.
We prepared 40 photos taken with mobile phones as avatars for the attacker's accounts. The photos were cut to 400x400 and converted into vectors by a trained model. The malware was published with the model and the vectors. The malware and attacker selected a trending topic once an hour in this experiment. Then, the attacker generated and posted the tweets, and the malware crawled related tweets 5 minutes later. In this experiment, the time was logged in a file when the attacker completed a hash collision, the malware crawled a batch of tweets, and the malware started and finished the comparisons. The original commands and the recovered commands were also logged in a file. Afterward, we used the logs to compare the post time and the retrieval time and determine the correctness of the recovered commands. 

(3) \textbf{Results}. 
We sent 47 commands using the 40 avatars. Due to frequent visits to Twitter trends, the selected topics are sometimes the same as the previous ones. Although it does not matter in real scenarios, we chose to wait for the next trending topic to evaluate the success rate of hash collisions more objectively. All commands in the experiments were received and parsed correctly by the seven hosts. During the tests, the attacker completed the tweet collection, tweets generation, and hash calculation in 13.8 s on average and reached a success rate of 90.28\% for hash collisions. After selecting a trending topic, the malware attempted to crawl 1 K tweets and usually obtained 800-900 non-repeated tweets (only original tweets were saved for retweeted tweets). The malware needed to crawl the avatars of the tweeters and calculate the distances to identify the attacker. Due to the different network and device conditions, the time this process required varied. The time costs for the malware to find the attacker are shown in Fig.~\ref{fig:mainres}. It takes 5 s to 4.45 min to find the attacker after crawling the tweets. During the experiments, some of our tweets received several ``likes'' from Twitter users. It shows the sentences generated by EDA did not cause anomalies and were acceptable. After the malware got the IPs, the attacker deleted the tweets.

\section{Evaluation}\label{sec:eva}
In this section, we evaluate the performance of different parts in DeepC2. \textbf{Environment:} The evaluation was performed on an Ubuntu 18.04 x64 virtual server with 1 GB ROM and 1 vCPU, and the code was implemented with Python 3.6.

\subsection{Tweets Generation} To test the efficiency of tweet generation for the attacker, we selected 79 trending topics from 4 randomly selected English-speaking areas around the world (San Francisco, London, Sydney, and Johannesburg). One-thousand tweets were crawled for each topic. Additionally, we cleaned the crawled tweets using the method in Sec.~\ref{sec:gentweets} and generated 50 new sentences using EDA for each remaining tweet. The trending topics may contain one or more words. With random deletion and random swap adopted in EDA, keywords in the topics may be deleted, or position changed in the newly generated sentences. The malware cannot find the attacker's accounts if the attacker posts sentences without exact keywords. Therefore, the number of sentences with accurate keywords and the quantity of all generated sentences were also recorded.

\begin{figure}
      \centering
      \includegraphics[width=\linewidth]{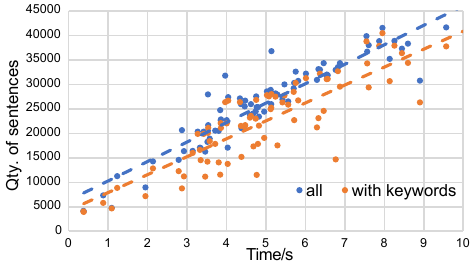}
      \caption{Efficiency of tweets generation}
      \label{fig:gentweets}
\end{figure}

In the 79 selected topics, 55 contained only one word, and 24 contained more than one word. With the percentage of words in each sentence to be changed set to 0.1, 89.54\% of the newly generated sentences contained accurate keywords for the 55 single-word topics, and 77.55\% contained accurate keywords for the 24 multi-word topics.
The time cost is linearly related to the number of the new sentences, as shown in Fig.~\ref{fig:gentweets}. As mentioned in Sec.~\ref{sec:gentweets}, EDA obtains 20 K to 45 K sentences in this experiment. According to the test, generating the sentences costs 3 to 10 seconds. It is acceptable for the attacker to prepare sentences for a hash collision.

\begin{figure}      
      \centering
      \includegraphics[width=\linewidth]{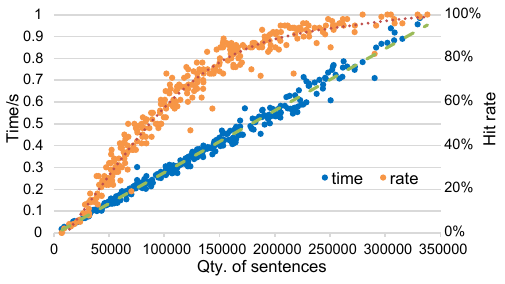}
      \caption{Time costs and hit rate of hash collisions}
      \label{fig:hasheff}
\end{figure}

\subsection{Hash Collision}\label{sec:hashexp} We used the sentences generated above to test the efficiency of hash collisions. To prepare different numbers of sentences, we followed the method in Sec.~\ref{sec:gentweets}, converted cases, and added punctuation at the end of the sentences. We got four batches of new sentences incrementally for each topic by adding two conversions once. We also collected 100 IP addresses as commands from a threat report~\cite{Group17}. We call a batch of sentences ``hit'' an IP if the batch succeeds in the hash collision. We used these new sentences and hashlib in Python 3.6.9 to calculate SHA-256 hashes on the virtual server with a single thread and recorded the time costs and hit rate of hash collisions with different quantities of sentences.

As shown in Fig.~\ref{fig:hasheff}, it took less than 1 second to calculate the hashes. In theory, 65,536 ($2^{16}$) sentences will hit an IP, which is ideal, as a hash collision is probabilistic. The experiment showed there should be at least 200 K sentences to obtain a 90\% hit rate and more than 330 K for a nearly 100\% hit rate. As mentioned in Sec.~\ref{sec:gentweets}, there are usually 140 K to more than 300 K sentences for the hash collision, and it will result in a hit rate above 75\%. During the experiments on Twitter, the attacker obtained an average of 219,335 sentences for hash collision and reached a hit rate of 90.28\%, which was also acceptable for practical purposes. Moreover, the attacker can crawl more trending tweets and generate more sentences in real scenarios.

\subsection{Avatar Recognition} To test the efficiency of avatar recognition by the malware, we used the 40 vectors above and 1,000 crawled avatars of size 400x400 to calculate the distances on the virtual server. The average time cost of extracting features from 1 K avatars and calculating 1 K distances was 11.92 s. It is also acceptable for the malware in such hardware conditions. In real scenarios, this process may take longer as the malware should crawl the avatars first, which varies due to different network conditions. Compared with the experiments on Twitter (Fig.~\ref{fig:mainres}), crawling the avatars is the most time-consuming process during the addressing.

\subsection{Crawling Tweets}\label{sec:crawling} In this experiment, the malware crawl 1 K tweets 5 minutes after the selection of the trending topic. In real scenarios, attackers can customize the waiting time, crawling volume, and frequency. In this part, we'll show how the attacker determines the appropriate parameters.

\begin{figure}
        \centering
        \includegraphics[width=0.9\linewidth]{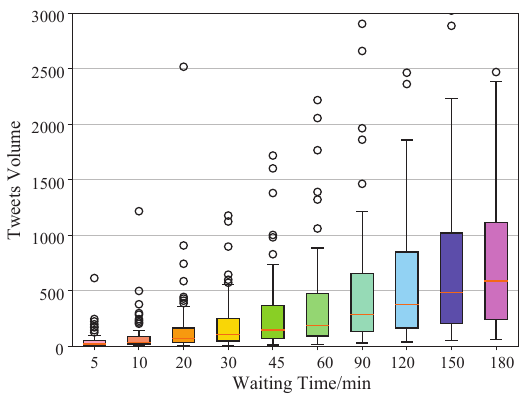}
        \caption{Crawling volume and frequency}
        \label{fig:waiting}
\end{figure}

We used the method in Sec.~\ref{sec:trends} to collect the trending topics. Then, we used the attacker's account to post tweets that contained the keywords. The malware started to find the attacker's account using the keywords after waiting for 5, 10, 20, 30, 45, 60, 90, 120, 150, and 180 minutes. The malware recorded how many tweets were crawled to find the attacker. We collected 56 groups of data. Fig.~\ref{fig:waiting} shows the relation between the crawled tweet volume and waiting time. After waiting 5 minutes, the malware found the attacker within 1 K tweets in all cases. After waiting for 1 hour, in 88\% of cases, the malware found the attacker within 1 K tweets and 98\% within 3 K tweets. After waiting for 3 hours, the malware could still find the attacker within 1 K tweets in 68\% of cases and within 3K tweets in 89\% of cases. As the waiting time is 5 minutes in the experiments on Twitter, it is appropriate to crawl 1,000 tweets.

The tweets may be more frequently updated if the attackers choose topics from larger cities such as New York and Los Angeles, and it may require the malware to crawl more tweets with the same waiting time. Additionally, if attackers choose top-ranked topics from the trending list, the malware also needs to crawl more tweets with the same waiting time. Moreover, it is also different if attackers choose to publish commands at midnight in the selected city. The parameters should be customized with different needs when applied in real scenarios.

\subsection{Security Analysis}\label{sec:sec_ana}

In this section, we discuss the security risks from the perspective of defenders. 

\textbf{Save and Reuse Avatars.} Although it is difficult to guess the avatars used by the attacker, the defender can monitor the behaviors of the compromised hosts to identify the attacker's accounts. The defender can reuse the attacker's avatars when the next appointed time arrives. They can also select a topic and post tweets that contain fake commands. This scenario will not work for the hosts always online because each avatar is used only once. However, hosts that go online after being offline and missing a command will recognize the defender's accounts as attackers and get an incorrect command. Therefore, authentication is recommended to ensure secure C\&C communication, as stated in Sec.~\ref{sec:gentweets}.

\begin{figure}
    \centering
    \includegraphics[scale=0.85]{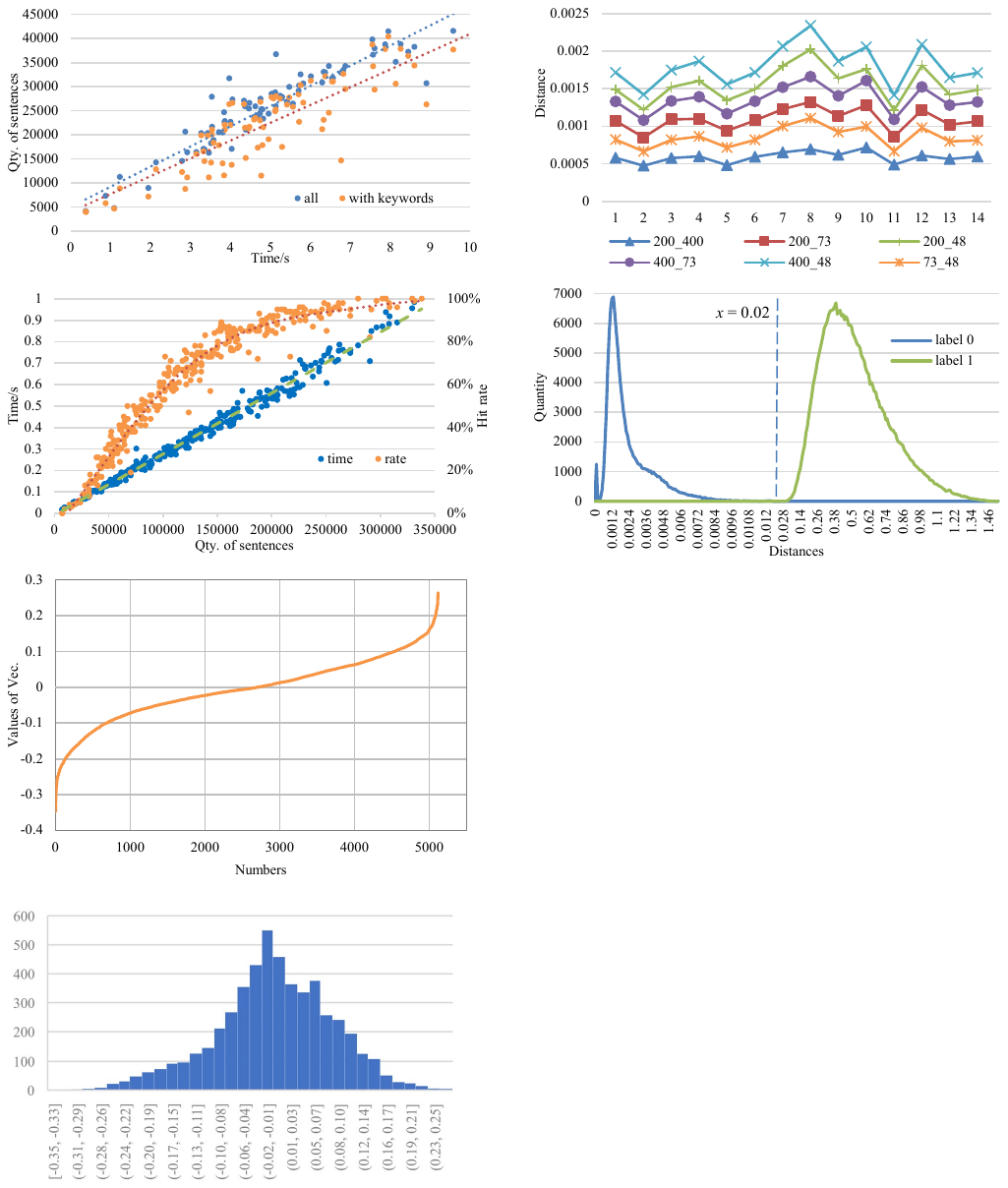}
    \caption{Values of vectors}
    \label{fig:vectors}
\end{figure}

\textbf{Collide an Avatar.} Defenders can try to collide an avatar. It sounds feasible but is hard practically. We analyzed the composition of the vectors. The 40 vectors in Sec.~\ref{sec:exp} contain 5,120 numbers. The numbers are sorted incrementally and put into the coordinate system, as shown in Fig.~\ref{fig:vectors}.
The numbers follow a normal distribution and constitute a continuous interval from -0.350 to 0.264.
Each vector value is taken from the interval, which is ample space and hard to enumerate or collide. It ensures the security of the attacker's avatars and vectors.

\begin{figure}
    \centering
    \includegraphics[scale=0.35]{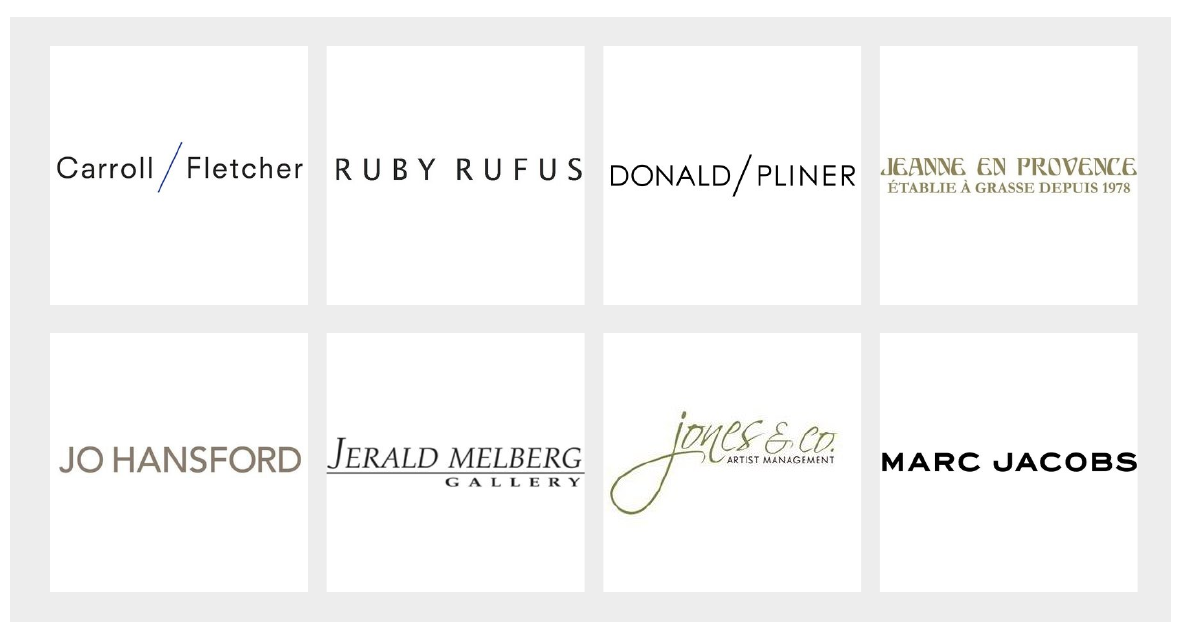}
    \caption{A group of avatars that have distances below 0.02}
    \label{fig:similar}
\end{figure}

However, we still attempted a collision for avatars. Using a trained model, we made more than 0.6 billion calculations on the distances between 115,887 pairs of crawled avatars. 2,050 avatar pairs have a distance below 0.02 (0.00031\%), of which 81 pairs are below 0.01 (0.000012\%). By analyzing these pictures, we found they share similar styles in that they all have a large solid color background, especially a white background (mainly logos) (see Fig.~\ref{fig:similar}). As avatars are prepared by attackers, they can avoid this type of picture. They can use colorful pictures taken by their cameras instead of pictures from the Internet.

\textbf{Train a GAN.} Defenders may train a GAN with saved avatars to generate similar images. Considering the computational costs, it is not feasible. As the avatars can be animals, plants, arts, etc., the training target is too divergent to be capable with GAN. Additionally, training a GAN needs numerous data, and the attacker's avatars are insufficient for building a training set.

\textbf{Train a Decoder.} Defenders have access to vectors and neural network models, so they can attempt to recover and derive a similar image from cheating the malware. CNN makes protection possible.
CNN learns abstract features from raw images. Each convolution layer generates a higher degree of abstraction from the previous layer. As layers deepen, much of the information in the original image is lost. This makes it difficult to recover the original image or derive a similar image based on the vectors.

We also simulated such an attack. We assume defenders treat the neural network as an encoder and build a corresponding decoder to generate related images. Defenders can also crawl avatars from Twitter and extract feature vectors using the model. The avatars and vectors make up the training data for the decoder. We trained such a decoder to generate numerous images from vectors and calculated the distance between the original image and the generated image. Due to the losses introduced by CNN and image conversion, the lowest distance we got is 0.0504, larger than the threshold. As avatars retrieved by the malware are not in the size of 128x128, more conversion and compression will be introduced to the images. It's also challenging to attack the C\&C in this way.

\textbf{Attack the Model.} Defenders can attack the neural network model to let the malware make incorrect decisions on attacker's accounts. There are some works on neural network Trojan attacks~\cite{LiuMALZW018}, which make this attack possible. As the target of this attack is a neural network model, it may affect some compromised hosts but does not influence the other hosts. Other unaffected hosts can still make correct decisions on the attacker's accounts.

\textbf{Generate Adversarial Samples.} As the model and feature vectors are known to defenders, it is a white-box non-targeted adversarial attack in this scenario~\cite{Qiu19}. Defenders can generate an adversarial sample to fool the model. Adversarial attacks aim at misclassifying the original target. Although CNN has 128 outputs, they don't represent 128 classes. Each output is a value in the feature vector. A slight perturbation of the value will result in a distance higher than the threshold. Therefore, it's not applicable to attack the C\&C in this way.


\section{Possible Countermeasures}\label{sec:discuss}
There are some ways to enhance the security of DeepC2, and we discuss them in Appendix~\ref{app:enhance}. In this section, we discuss the possible countermeasures.

\textbf{Behavior Analysis.}
Traditional malware detection methods such as behavior analysis and traffic analysis can be applied to detect the malware~\cite{botminer08}. There are periodic behaviors of the malware. They need to visit Twitter Trends periodically. After selecting a trending topic, they need to crawl tweets and avatars to find attackers. This series of operations can make up a behavioral pattern. In addition, the periodic net flow is also a noticeable feature. 

\textbf{Collaboration.} In this scenario, it is recommended that security analysts share the malware samples to the communities and the related OSNs once they appear so that every party can contribute to the mitigating works. OSNs can detect attackers in real-time by running the samples and actively monitoring activities related to the malware and attackers. They can calculate the distances between the uploaded avatars and vectors and block the attackers as soon as the corresponding avatars are detected. This may need a large-scale calculation but is an effective way to mitigate this attack. Meanwhile, OSNs can also help to trace the attackers behind the accounts. Therefore, we believe the cooperation between OSNs and security communities is essential to mitigate this attack.

\textbf{Improvement on OSNs.} There are many ways to utilize OSNs, so OSNs should take measures to avoid abuse.
The attackers should maintain some Twitter accounts. The accounts can be stolen from ordinary users, registered in bulk using automated programs~\cite{TwitterRegBulk2021}, or brought from underground markets~\cite{TwitterBuy2021}.
Therefore, we suggest OSNs apply more complex human-machine verification during the registration and manage the misbehaved social bots under the terms of services (ToS).
Cracking down on underground account transactions is also necessary. While working on this work, we found some websites selling Twitter accounts in bulk. We cannot predict how they got the accounts and how the buyers use the accounts. Since it violates Twitter ToS~\cite{Twitter20}, related parties should limit illegal account transactions. We have reported it to Twitter.


As AI can be used to launch cyberattacks, security vendors should also consider the malicious use of AI so that the attacks can be detected when they are applied in real scenarios in the future.

\section{Conclusion}\label{sec:conclusion}

This paper discussed a novel covert command and control scenario, DeepC2, on OSNs by introducing AI technologies. By utilizing the poor explainability of neural network models, the addressing process can be concealed in AI models rather than exposed as reversible hard-coding. For issuing commands covertly, we use easy data augmentation and hash collision to generate contextual and readable command-embedded tweets to avoid abnormal content on OSNs. We conduct experiments on Twitter to show the feasibility and efficiency. Furthermore, we analyze the security of the avatars. We also discussed possible countermeasures to mitigate this kind of attack.

AI is also capable of attacks. With the popularity of AI, AI-powered attacks will emerge and bring new challenges to cybersecurity. Cyberattacks and defense are interdependent. We believe countermeasures against AI attacks will be applied in future computer systems, and protection for computer systems will be more intelligent. We hope the proposed scenario will contribute to future protection efforts.

\bibliographystyle{IEEEtran}
\bibliography{sample-base}

\begin{thebibliography}{10}
\providecommand{\url}[1]{#1}
\csname url@samestyle\endcsname
\providecommand{\newblock}{\relax}
\providecommand{\bibinfo}[2]{#2}
\providecommand{\BIBentrySTDinterwordspacing}{\spaceskip=0pt\relax}
\providecommand{\BIBentryALTinterwordstretchfactor}{4}
\providecommand{\BIBentryALTinterwordspacing}{\spaceskip=\fontdimen2\font plus
\BIBentryALTinterwordstretchfactor\fontdimen3\font minus
  \fontdimen4\font\relax}
\providecommand{\BIBforeignlanguage}[2]{{%
\expandafter\ifx\csname l@#1\endcsname\relax
\typeout{** WARNING: IEEEtran.bst: No hyphenation pattern has been}%
\typeout{** loaded for the language `#1'. Using the pattern for}%
\typeout{** the default language instead.}%
\else
\language=\csname l@#1\endcsname
\fi
#2}}
\providecommand{\BIBdecl}{\relax}
\BIBdecl

\bibitem{Bailey09}
M.~{Bailey}, E.~{Cooke}, F.~{Jahanian}, Y.~{Xu}, and M.~{Karir}, ``A survey of
  botnet technology and defenses,'' in \emph{2009 Cybersecurity Applications
  Technology Conference for Homeland Security}, 2009, pp. 299--304.

\bibitem{Yin18}
J.~Yin, H.~Lv, F.~Zhang, Z.~Tian, and X.~Cui, ``Study on advanced botnet based
  on publicly available resources,'' in \emph{Information and Communications
  Security}.\hskip 1em plus 0.5em minus 0.4em\relax Cham: Springer
  International Publishing, 2018, pp. 57--74.

\bibitem{FireEye15}
FireEye, ``Uncovering a malware backdoor that uses twitter,'' FireEye, Tech.
  Rep., 2015.

\bibitem{Stefanko18}
\BIBentryALTinterwordspacing
L.~Stefanko. (2018, June) New telegram‑abusing android rat discovered in the
  wild. [Online]. Available:
  \url{https://www.welivesecurity.com/2018/06/18/new-telegram-abusing-android-rat/}
\BIBentrySTDinterwordspacing

\bibitem{Faou20}
M.~Faou, ``From agent.btz to comrat v4: A ten-year journey,'' ESET, Tech. Rep.,
  May 2020.

\bibitem{upd4t309}
\BIBentryALTinterwordspacing
A.~{Moscaritolo}. (2009, Aug) Twitter used as botnet command-and-control hub.
  [Online]. Available:
  \url{https://www.itnews.com.au/news/twitter-used-as-botnet-command-and-control-hub-153144}
\BIBentrySTDinterwordspacing

\bibitem{singh2012social}
A.~Singh, ``Social networking for botnet command and control,'' 2012.

\bibitem{sebastian2014framework}
S.~Sebastian, S.~Ayyappan, and P.~Vinod, ``Framework for design of graybot in
  social network,'' in \emph{2014 International Conference on Advances in
  Computing, Communications and Informatics (ICACCI)}.\hskip 1em plus 0.5em
  minus 0.4em\relax IEEE, 2014, pp. 2331--2336.

\bibitem{F-Secure15}
F-Secure, ``The dukes: 7 years of russian cyberespionage,'' F-Secure, Tech.
  Rep., 2015.

\bibitem{Pantic15}
N.~Pantic and M.~I. Husain, ``Covert botnet command and control using
  twitter,'' in \emph{Proceedings of the 31st Annual Computer Security
  Applications Conference}, ser. ACSAC 2015.\hskip 1em plus 0.5em minus
  0.4em\relax ACM, 2015, p. 171–180.

\bibitem{ROKRAT17}
\BIBentryALTinterwordspacing
W.~Mercer, P.~Rascagneres, and M.~Molyett. (2017, Apr) Introducing rokrat.
  [Online]. Available:
  \url{https://blog.talosintelligence.com/2017/04/introducing-rokrat.html}
\BIBentrySTDinterwordspacing

\bibitem{PlugX17}
\BIBentryALTinterwordspacing
T.~{Lancaster} and E.~{Idrizovic}. (2017, June) Paranoid plugx. [Online].
  Available: \url{https://unit42.paloaltonetworks.com/unit42-paranoid-plugx}
\BIBentrySTDinterwordspacing

\bibitem{Comnie18}
\BIBentryALTinterwordspacing
J.~{Grunzweig}. (2018, Jan) Comnie continues to target organizations in east
  asia. [Online]. Available:
  \url{https://unit42.paloaltonetworks.com/unit42-comnie-continues-target-organizations-east-asia/}
\BIBentrySTDinterwordspacing

\bibitem{DarkHydrus19}
\BIBentryALTinterwordspacing
R.~{Falcone} and B.~{Lee}. (2019, Jan) Darkhydrus delivers new trojan that can
  use google drive for c2 communications. [Online]. Available:
  \url{https://unit42.paloaltonetworks.com/darkhydrus-delivers-new-trojan-that-can-use-google-drive-for-c2-communications/}
\BIBentrySTDinterwordspacing

\bibitem{xai20}
\BIBentryALTinterwordspacing
F.~Lecue, K.~Gade, S.~C. Geyik, K.~Kenthapadi, V.~Mithal, A.~Taly, R.~Guidotti,
  and P.~Minervini. (2020, Feb) Explainable ai: Foundations, industrial
  applications, practical challenges, and lessons learned. [Online]. Available:
  \url{https://xaitutorial2020.github.io/}
\BIBentrySTDinterwordspacing

\bibitem{DGA16}
D.~Plohmann, K.~Yakdan, M.~Klatt, J.~Bader, and E.~Gerhards-Padilla, ``A
  comprehensive measurement study of domain generating malware,'' in \emph{25th
  {USENIX} Security Symposium}.\hskip 1em plus 0.5em minus 0.4em\relax Austin,
  TX: {USENIX} Association, Aug. 2016, pp. 263--278.

\bibitem{pony2021}
T.~Taniguchi, H.~Griffioen, and C.~Doerr, ``Analysis and takeover of the
  bitcoin-coordinated pony malware,'' in \emph{Proceedings of the 2021 ACM Asia
  Conference on Computer and Communications Security}.\hskip 1em plus 0.5em
  minus 0.4em\relax ACM, 2021, p. 916–930.

\bibitem{Shishir11}
S.~Nagaraja, A.~Houmansadr, P.~Piyawongwisal, V.~Singh, P.~Agarwal, and
  N.~Borisov, ``Stegobot: A covert social network botnet,'' in
  \emph{Information Hiding}.\hskip 1em plus 0.5em minus 0.4em\relax Berlin,
  Heidelberg: Springer Berlin Heidelberg, 2011, pp. 299--313.

\bibitem{Kwak21}
M.~Kwak and Y.~Cho, ``A novel video steganography-based botnet communication
  model in telegram sns messenger,'' \emph{Symmetry}, vol.~13, no.~1, 2021.

\bibitem{Wei19}
J.~W. Wei and K.~Zou, ``{EDA:} easy data augmentation techniques for boosting
  performance on text classification tasks,'' in \emph{Proceedings of the 2019
  Conference on Empirical Methods in Natural Language Processing and the 9th
  International Joint Conference on Natural Language Processing, {EMNLP-IJCNLP}
  2019, Hong Kong, China}, 2019, pp. 6381--6387.

\bibitem{MITREattack20}
\BIBentryALTinterwordspacing
ATT\&CK. (2020). [Online]. Available:
  \url{https://twitter.com/MITREattack/status/1267815021301633024}
\BIBentrySTDinterwordspacing

\bibitem{RigakiG18}
M.~{Rigaki} and S.~{Garcia}, ``Bringing a {GAN} to a knife-fight: Adapting
  malware communication to avoid detection,'' in \emph{2018 {IEEE} Security and
  Privacy Workshops, {SP} Workshops 2018, San Francisco, CA, USA}.\hskip 1em
  plus 0.5em minus 0.4em\relax {IEEE} Computer Society, 2018, pp. 70--75.

\bibitem{Liu20stego}
T.~Liu, Z.~Liu, Q.~Liu, W.~Wen, W.~Xu, and M.~Li, ``Stegonet: Turn deep neural
  network into a stegomalware,'' in \emph{Annual Computer Security Applications
  Conference}, ser. ACSAC '20.\hskip 1em plus 0.5em minus 0.4em\relax New York,
  NY, USA: Association for Computing Machinery, 2020, p. 928–938.

\bibitem{Wang2021EvilModel}
Z.~Wang, C.~Liu, and X.~Cui, ``{EvilModel: Hiding Malware Inside of Neural
  Network Models},'' in \emph{{IEEE} Symposium on Computers and Communications,
  {ISCC} 2021, Athens, Greece, September 5-8, 2021}.\hskip 1em plus 0.5em minus
  0.4em\relax {IEEE}, 2021, pp. 1--7.

\bibitem{Wang2022EvilModel2}
Z.~Wang, C.~Liu, X.~Cui, J.~Yin, and X.~Wang, ``{EvilModel} 2.0: Bringing
  neural network models into malware attacks,'' \emph{Computers \& Security},
  vol. 120, p. 102807, 2022.

\bibitem{Kirat18}
D.~Kirat, J.~Jang, and M.~P. Stoecklin, ``Deeplocker - concealing targeted
  attacks with ai locksmithing,'' IBM Research, Tech. Rep., 2018.

\bibitem{HuT17}
\BIBentryALTinterwordspacing
W.~Hu and Y.~Tan, ``Generating adversarial malware examples for black-box
  attacks based on {GAN},'' \emph{CoRR}, vol. abs/1702.05983, 2017. [Online].
  Available: \url{http://arxiv.org/abs/1702.05983}
\BIBentrySTDinterwordspacing

\bibitem{EvadePE18}
\BIBentryALTinterwordspacing
H.~S. Anderson, A.~Kharkar, B.~Filar, D.~Evans, and P.~Roth, ``Learning to
  evade static {PE} machine learning malware models via reinforcement
  learning,'' \emph{CoRR}, vol. abs/1801.08917, 2018. [Online]. Available:
  \url{http://arxiv.org/abs/1801.08917}
\BIBentrySTDinterwordspacing

\bibitem{Wang21}
J.~Wang, Q.~Liu, D.~Wu, Y.~Dong, and X.~Cui, ``Crafting adversarial example to
  bypass flow-{\&}ml- based botnet detector via {RL},'' in \emph{{RAID} '21:
  24th International Symposium on Research in Attacks, Intrusions and Defenses,
  San Sebastian, Spain, October 6-8, 2021}.\hskip 1em plus 0.5em minus
  0.4em\relax {ACM}, 2021, pp. 193--204.

\bibitem{google_nnmodel22}
S.~Chatterjee and P.~Zielinski, ``On the generalization mystery in deep
  learning,'' \emph{CoRR}, vol. abs/2203.10036, 2022.

\bibitem{ImageHash2021}
\BIBentryALTinterwordspacing
J.~Buchner. (2020) Imagehash-pypi. [Online]. Available:
  \url{https://pypi.org/project/ImageHash/}
\BIBentrySTDinterwordspacing

\bibitem{ImageHashAAAI}
R.~Xia, Y.~Pan, H.~Lai, C.~Liu, and S.~Yan, ``Supervised hashing for image
  retrieval via image representation learning,'' in \emph{Proceedings of the
  Twenty-Eighth {AAAI} Conference on Artificial Intelligence}.\hskip 1em plus
  0.5em minus 0.4em\relax {AAAI} Press, 2014, pp. 2156--2162.

\bibitem{sbox1985}
A.~F. Webster and S.~E. Tavares, ``On the design of s-boxes,'' in
  \emph{Advances in Cryptology --- CRYPTO '85 Proceedings}.\hskip 1em plus
  0.5em minus 0.4em\relax Berlin, Heidelberg: Springer, 1986, pp. 523--534.

\bibitem{snn94}
J.~Bromley, J.~W. Bentz, L.~Bottou, I.~Guyon, Y.~LeCun, C.~Moore,
  E.~S{\"{a}}ckinger, and R.~Shah, ``Signature verification using {A}
  ``siamese'' time delay neural network,'' \emph{Int. J. Pattern Recognit.
  Artif. Intell.}, vol.~7, no.~4, pp. 669--688, 1993.

\bibitem{LuCun89}
Y.~LeCun, B.~Boser, J.~S. Denker, D.~Henderson, R.~E. Howard, W.~Hubbard, and
  L.~D. Jackel, ``Backpropagation applied to handwritten zip code
  recognition,'' \emph{Neural Computation}, vol.~1, no.~4, pp. 541--551, 1989.

\bibitem{LeCun06}
R.~{Hadsell}, S.~{Chopra}, and Y.~{LeCun}, ``Dimensionality reduction by
  learning an invariant mapping,'' in \emph{2006 IEEE Computer Society
  Conference on Computer Vision and Pattern Recognition (CVPR'06)}, vol.~2,
  2006, pp. 1735--1742.

\bibitem{Group17}
Group‑IB, ``Lazarus arisen: Architecture, techniques and attribution,''
  Group‑IB, Tech. Rep., 2017.

\bibitem{LiuMALZW018}
Y.~{Liu}, S.~{Ma}, Y.~{Aafer}, W.~{Lee}, J.~{Zhai}, W.~{Wang}, and X.~{Zhang},
  ``Trojaning attack on neural networks,'' in \emph{25th Annual Network and
  Distributed System Security Symposium, {NDSS} 2018}.

\bibitem{Qiu19}
S.~Qiu, Q.~Liu, S.~Zhou, and C.~Wu, ``Review of artificial intelligence
  adversarial attack and defense technologies,'' \emph{Applied Sciences},
  vol.~9, no.~5, p. 909, 2019.

\bibitem{botminer08}
G.~Gu, R.~Perdisci, J.~Zhang, and W.~Lee, ``Botminer: Clustering analysis of
  network traffic for protocol- and structure-independent botnet detection,''
  in \emph{Proceedings of the 17th {USENIX} Security Symposium}.\hskip 1em plus
  0.5em minus 0.4em\relax {USENIX} Association, 2008, pp. 139--154.

\bibitem{TwitterRegBulk2021}
\BIBentryALTinterwordspacing
Quora. (2020) How can i create bulk twitter accounts automatically? [Online].
  Available:
  \url{https://www.quora.com/How-can-I-create-bulk-Twitter-accounts-automatically}
\BIBentrySTDinterwordspacing

\bibitem{TwitterBuy2021}
\BIBentryALTinterwordspacing
Google. (2021) Google search. [Online]. Available:
  \url{https://www.google.com/search?q=buy+twitter+accounts}
\BIBentrySTDinterwordspacing

\bibitem{Twitter20}
\BIBentryALTinterwordspacing
Twitter. (2020) Twitter terms of service. [Online]. Available:
  \url{https://twitter.com/en/tos}
\BIBentrySTDinterwordspacing

\bibitem{ferrara2016rise}
E.~Ferrara, O.~Varol, C.~Davis, F.~Menczer, and A.~Flammini, ``The rise of
  social bots,'' \emph{Communications of the ACM}, vol.~59, no.~7, pp. 96--104,
  2016.

\end{thebibliography}

\appendix

\subsection{Selection on Avatar Size}\label{app:size}

Twitter provides 4 different sizes of avatars: 48x48, 73x73, 200x200 and 400x400 (see Table~\ref{tab:avatars}). Links to these avatars of the same user are similar, and the only difference lies on the sizes. We need to use avatars from the same user to make up the training set with label 0. There are 6 combinations of the 4 sizes. To better serve the training set, an avatar combination should have a shorter distance as they come from the same user. Therefore, 256 avatars from 64 users are selected to calculate the distances using a CNN modified on LeNet (see Table~\ref{tab:lenet}).

\begin{table}[]
    \centering
    \caption{Links for different size of avatars of the same user}
    \begin{tabular}{c|p{6cm}}
      \hline
         \textbf{Size}&\textbf{Link}\\
      \hline
      400x400&\href{https://pbs.twimg.com/profile_images/1354479896072884225/AaUbc7ao_400x400.jpg}{https://pbs.twimg.com/profile\_images/13544798 96072884225/AaUbc7ao\_400x400.jpg}\\
      200x200&\href{https://pbs.twimg.com/profile_images/1354479896072884225/AaUbc7ao_200x200.jpg}{https://pbs.twimg.com/profile\_images/13544798 96072884225/AaUbc7ao\_200x200.jpg}\\
      73x73&\href{https://pbs.twimg.com/profile_images/1354479896072884225/AaUbc7ao_bigger.jpg}{https://pbs.twimg.com/profile\_images/13544798 96072884225/AaUbc7ao\_bigger.jpg}\\
      48x48&\href{https://pbs.twimg.com/profile_images/1354479896072884225/AaUbc7ao_normal.jpg}{https://pbs.twimg.com/profile\_images/13544798 96072884225/AaUbc7ao\_normal.jpg}\\
      \hline
    \end{tabular}
    \label{tab:avatars}
\end{table}

\begin{table}[]
    \centering
    \caption{Architecture of CNN}
    \label{tab:lenet}
    \begin{tabular}{ccccc}
      \toprule
        \textbf{layer}&\textbf{size-in}&\textbf{size-out}&\textbf{kernel}&\textbf{param}\\
      \midrule
        conv1&32×32×3&28×28×6&5×5×6, 1&0.5K\\
        ReLU&&&&\\
        pool1&28×28×6&14×14×6&2×2×1, 2&0\\ \hline
        conv2&14×14×6&10×10×16&5×5×16, 1&2.4K\\
        Sigmoid&&&&\\
        pool2&10×10×16&5×5×16&2×2×1, 2&0\\ \hline
        fc1&1×400×1&1×256×1& &102.7K\\
        Sigmoid&&&&\\ \hline
        fc2&1×256×1&1×192×1& &49.3K\\
        ReLU&&&&\\ \hline
        output&1×192×1&1×128×1& &24.7K\\ \midrule
        \textbf{total}& & & &179.6K\\
      \bottomrule
    \end{tabular}
\end{table}

A total of 256 vectors are extracted from the avatars by the CNN with random weights. For each user, there are 4 vectors that represent 4 different sizes of avatars. Then, Euclidean distances of different vectors from the same user are calculated. Now there are 6 distances for each user. For simplicity, we use $D_{n\_m}$ to represent the distance between avatars of size n×n and m×m. The experiment was repeated 14 times with 14 different random weights for the CNN.

\begin{figure}
    \centering
    \includegraphics[scale=0.9]{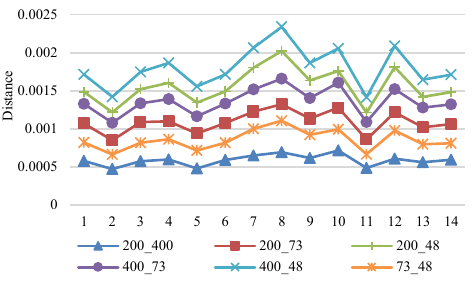}
    \caption{Average Distances for Different Avatar Sizes}
    \label{fig:six}
\end{figure}

Fig.~\ref{fig:six} shows the 14 average distances of the 6 size combinations. This indicates that when one avatar is fixed in size, a smaller size gap gains smaller distances (e.g. [$D_{400\_73}$, $D_{200\_73}$, $D_{73\_48}$]), and avatars with a larger size gain smaller distances (e.g. [$D_{400\_200}$, $D_{73\_48}$]). Therefore, avatars with sizes of 200x200 and 400x400 are used in the training set because they have the shortest distance of all the combinations, and avatars with sizes of 400x400 are used by bots to find botmasters.

\begin{figure}
    \centering
    \includegraphics[scale=0.9]{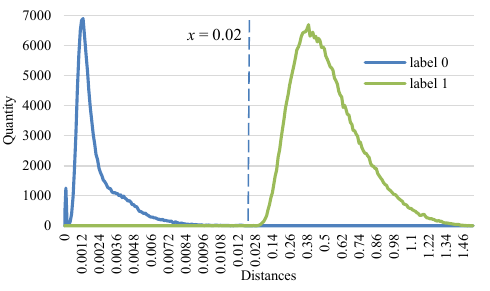}
    \caption{Threshold for Distance}
    \label{fig:threshold}
\end{figure}

\subsection{Threshold for Distance}\label{app:threshold}

A threshold is needed to determine whether two avatars share the same source. We use a trained model to calculate the distances on the validation set, which contains 115,887 pairs with label 0 and 347,657 pairs with label 1. We record the distances of every comparison, sort them by value and label, and count their frequencies to learn the boundary between the ``same'' avatars and different avatars. As shown in Fig.~\ref{fig:threshold}, the distances of all pairs with label 1 and only 4 pairs with label 0 are larger than 0.02, and the remaining pairs with label 0 are less than 0.02. It shows that 0.02 is a proper threshold for the determination. In real scenarios, attackers can choose a threshold less than 0.02, as the undistributed avatars and distances are within the authority of attackers.

\subsection{Enhancement}\label{app:enhance}

As proof of concept, the parameters in this work are conservative. There are ways to enhance the security of DeepC2.

In the model's design, the vectors can be longer than 128, making analysis and collisions for avatars even more difficult. The threshold of distances can also be lower than 0.02, as the undistributed avatars and the distances are within the authority of attackers. They can balance efficiency and accuracy according to the needs. Additionally, more losses can be introduced during the processing of avatars, like compression, deformation, format conversion, etc., making it harder to recover the avatars.

For addressing, the attacker can select more topics. Attackers can publish commands on the topics, and the malware can choose one randomly to find attackers. Attackers can also use other fields in OSNs to convey customized content. For instance, attackers could comment on a tweet, and the malware would identify and obtain commands from attackers' profiles. Other platforms, like Weibo and Tumblr, can also be utilized.

As stated before, attackers should maintain some accounts to publish different commands. To reduce the specious behaviors of accounts, attackers can maintain them by imitating normal users or social bots~\cite{ferrara2016rise}. This work can be done manually or automatically~\cite{TwitterRegBulk2021}. When attackers need to publish a command, attackers can select one account and maintain other accounts as usual.

\end{document}